\documentstyle[12pt]{article}

\newcommand{\pr}{\hspace{\parindent}}
\newcommand{\bea}{\begin{eqnarray}}
\newcommand{\eea}{\end{eqnarray}}
\newcommand{\beq}{\begin{equation}}
\newcommand{\eeq}{\end{equation}}

\def\msbar{\ifmmode{\overline{\rm MS}} \else{$\overline{\rm MS}$} \fi}
\def\drbar{\ifmmode{\overline{\rm DR}} \else{$\overline{\rm DR}$} \fi}
\def\to{\rightarrow}
\def\su{\ifmmode{\tilde{u}} \else{$\tilde{u}$} \fi}
\def\sd{\ifmmode{\tilde{d}} \else{$\tilde{d}$} \fi}
\def\sq{\ifmmode{\tilde{q}} \else{$\tilde{q}$} \fi}
\def\sg{\ifmmode{\tilde{g}} \else{$\tilde{g}$} \fi}
\def\snu{\ifmmode{\tilde{\nu}} \else{$\tilde{\nu}$} \fi}
\def\se{\ifmmode{\tilde{e}} \else{$\tilde{e}$} \fi}
\def\smu{\ifmmode{\tilde{\mu}} \else{$\tilde{\mu}$} \fi}
\def\sfp{\ifmmode{\tilde{f}_{1L}} \else{$\tilde{f}_{1L}$} \fi}
\def\sfm{\ifmmode{\tilde{f}_{2L}} \else{$\tilde{f}_{2L}$} \fi}
\def\simgt{\rlap{\lower 3.5 pt \hbox{$\mathchar \sim$}}%
           \raise 1pt \hbox {$>$}}
\def\simlt{\rlap{\lower 3.5 pt \hbox{$\mathchar \sim$}}%
           \raise 1pt \hbox {$<$}}

\begin{document}

\hfill\vbox{\baselineskip14pt
            \hbox{KEK-TH-446}
            \hbox{KEK Preprint 95-94}
            \hbox{hep-ph/9507419}
            \hbox{July 1995}}
\vspace{10mm}

\baselineskip22pt
\begin{center}
\Large  Supersymmetric contribution to the quark-lepton universality
violation in charged currents
\end{center}
\vspace{10mm}

\begin{center}
\large  K.~Hagiwara, S.~Matsumoto and Y.~Yamada
\end{center}
\vspace{0mm}

\begin{center}
\begin{tabular}{l}
Theory Group, KEK, Tsukuba, Ibaraki 305, Japan\\
\end{tabular}
\end{center}

\vspace{20mm}
\begin{center}
\large Abstract
\end{center}
\begin{center}
\begin{minipage}{13cm}
\baselineskip=22pt
\noindent
The supersymmetric one-loop contribution to the
quark-lepton universality violation in the low-energy charged current weak
interactions is studied.
It is shown that the recent experimental data with the
1-$\sigma$ deviation from the universality may be explained by the
existence of the light sleptons ($<220$GeV) and relatively light
chargino and neutralinos ($<600$GeV) with significant gaugino components.
The sign of the universality violation is briefly discussed.
\end{minipage}
\end{center}
\vfill
\newpage

\baselineskip=22pt
\normalsize
\setcounter{footnote}{0}

The tree-level universality of the charged current weak interactions is one
of the important consequences of the SU(2)$_L$ gauge symmetry of the
fundamental theory.
The universality between quarks and leptons is expressed
as the unitarity of the Cabibbo-Kobayashi-Maskawa (CKM) matrix, for example,
\beq
|V_{ud}|^2+|V_{us}|^2+|V_{ub}|^2=1. \label{1}
\eeq
Experimentally, these CKM matrix elements are extracted from the ratios of
the amplitude of the
semileptonic hadron decays to that of the muon decay.
In this paper, we adopt the values quoted in refs.\cite{towner,sirlin2,rpp},
\beq
|V_{ud}|=0.9745\pm0.0007[1,2],\;\; |V_{us}|=0.2205\pm0.0018[3],\;\;
|V_{ub}|=0.003\pm0.001[3]. \label{2}
\eeq
Their squared sum is then
\beq
|V_{ud}|^2+|V_{us}|^2+|V_{ub}|^2-1=-0.0017\pm0.0015. \label{3}
\eeq
The universality is violated at 1-$\sigma$ level.

In general, the universality (\ref{1}) can be modified by
the process-dependent radiative corrections, due to the spontaneous
breaking of the SU(2)$_L$ symmetry.
The data (\ref{2}), which are obtained after subtracting the known Standard
Model corrections \cite{sirlin,leut,towner,sirlin2},
lead to a small deviation from the quark-lepton universality,
although it is only at the 1-$\sigma$ level. This may suggest
a signal for physics beyond the Standard Model (SM).

In this paper, we study the possibility that the universality violation
(\ref{3}) is due to the radiative correction of the unknown new particles
in the minimal supersymmetric (SUSY) standard model, MSSM \cite{nilles,mssm},
which is one of the most interesting and promising
extensions of the SM.
We study the one-loop contribution of the SUSY particles
to the quark-lepton universality violation in low-energy charged currents.

There are preceding works \cite{barbieri1,barbieri2} on this subject.
In ref.\cite{barbieri1}, they studied only cases with very light
sfermions which are now experimentally excluded,
within the constraints on the MSSM parameter, $\tan\beta=1$.
In ref.\cite{barbieri2},
the squark-loop contributions were ignored and only the range of the
universality violation under the LEP-I constraints were shown.
We refine these works by extending the analysis to cover the whole
parameter space of the MSSM,
and obtain constraints on SUSY particle masses from the 1-$\sigma$
universality violation (\ref{3}).
In addition, implications of the sign of the universality violation
to the SUSY model parameters are briefly discussed.

We first fix our framework, following the formalism given in
ref.\cite{hhkm}. We study the decay $f_2\to f_1\ell^-\bar{\nu}$
where $f=(f_1, f_2)$ is a SU(2)$_L$ fermion doublet,
comparing the following two cases;
muon decay $(f_1=\nu_{\mu}, f_2=\mu^-)$ and
semileptonic hadron decays $(f_1=u, f_2=d, s, b)$.
At the tree level, their decay amplitudes are identical
up to the CKM matrix element, and are
expressed in terms of the bare Fermi constant $G_0=g^2/4\sqrt{2}m_W^2$.
At the one-loop level, however, the decay amplitudes receive process-dependent
radiative corrections as well as process-independent (oblique) ones.
Following ref.\cite{hhkm}, the corrected decay amplitudes
of $f_2\to f_1\ell^-\bar{\nu}$ are expressed as
\beq
G_{f}=\frac{\bar{g}_W^2(0)+g^2\bar{\delta}_{Gf}}
{4\sqrt{2}m_W^2}. \label{4}
\eeq
The effective coupling $\bar{g}_W^2(0)$ represents the correction to the
W-boson propagator and does not lead to the universality violation.
The process-dependent term $\bar{\delta}_{Gf}$, which represents the
vertex and box corrections, gives the universality violation.
In the MSSM,
$\bar{\delta}_{Gf}=\bar{\delta}_{Gf}({\rm SM})+\delta_{Gf}({\rm SUSY})$,
where $\bar{\delta}_{Gf}({\rm SM})$ is the gauge vector
loop contribution and $\delta_{Gf}({\rm SUSY})$ is the SUSY loop
contribution.

Here we address the definition of the CKM matrix elements in eq.(\ref{2}).
In testing the universality of the charged current interactions in the SM,
the process-dependent corrections $\bar{\delta}_{Gf}({\rm SM})$
should be accounted for.
In fact, the value of $V_{ud}$ quoted in eq.(\ref{2})
is extracted from the nuclear $\beta$-decay transition rates
after subtracting the SM radiative corrections and correcting for nuclear
form factor effects \cite{towner,sirlin2,rpp,sirlin,towner2}.
Without the SM corrections, the sum (\ref{3}) would be
about 1.04, showing more than 10-$\sigma$ deviation
from the universality \cite{sirlin2}.
We note here that the theoretical uncertainties in estimating
the nuclear form factor effects dominate the error assigned
to $V_{ud}$ \cite{towner,sirlin2,rpp,sirlin,towner2}.
The effect of $\bar{\delta}_{Gf}({\rm SM})$ has also been subtracted
\cite{leut} in extracting the $V_{us}$ matrix element in eq.(\ref{2}).
The SM correction for $V_{ub}$ has no significant effect on our analysis.
Therefore, in our study, the 1-$\sigma$ deviation of eq.(\ref{3})
from unity is interpreted as an effect of the non-universal
corrections by the SUSY particle loops, $\delta_{Gf}({\rm SUSY})$.

We show the analytic form of
$\delta_{Gf}({\rm SUSY})$, basically following the
notation of ref.\cite{gh}.
$\delta_{Gf}({\rm SUSY})$ comes from the loops with left-handed
sfermions ($\snu_e$, $\se$, $\tilde{f}_1$, $\tilde{f}_2$)$_L$,
charginos $\tilde{C}_j(j=1,2)$, neutralinos $\tilde{N}_i(i=1-4)$
and a gluino \sg.
It is expressed as a sum
$\delta_f^{(v)}+\delta_{\ell}^{(v)}+\delta_f^{(b)}$, where we use the
abbreviations $\delta_f\equiv\delta_{Gf}({\rm SUSY})$ etc.
The correction $\delta_f^{(v)}$ to the $W^+\bar{f}_1f_2$ vertex is
\bea
\lefteqn{(4\pi)^2\delta_f^{(v)}}
\nonumber \\
&&=\frac{g^2}{2}\sum_i[|V_{i1}|^2B_1(\tilde{C}_i,\sfp)+|U_{i1}|^2
B_1(\tilde{C}_i,\sfm)]
\nonumber\\
&&+g_Z^2\sum_i[|g_{NiL}^{(f_2)}|^2B_1(\tilde{N}_i,\sfm)+
|g_{NiL}^{(f_1)}|^2B_1(\tilde{N}_i,\sfp)]\nonumber\\
&&+2g_Z^2\sum_ig_{NiL}^{(f_1)}g_{NiL}^{(f_2)*}
2C_{24}(\sfm,\tilde{N}_i,\sfp)\nonumber\\
&&-2gg_Z\sum_{i,j}U_{j1}g_{NiL}^{(f_2)*}
\left[ l_{ij}^{0-}(2C_{24}(\tilde{N}_i,\sfm,\tilde{C}_j)-\frac{1}{2})
+r_{ij}^{0-}m_{\tilde{C}_j}m_{\tilde{N}_i}
(-C_0(\tilde{N}_i,\sfm,\tilde{C}_j))\right] \nonumber\\
&&+2gg_Z\sum_{i,j}V_{j1}^*g_{NiL}^{(f_1)}
\left[ r_{ij}^{0-}(2C_{24}(\tilde{C}_j,\sfp, \tilde{N}_i)-\frac{1}{2})
+l_{ij}^{0-}m_{\tilde{C}_j}m_{\tilde{N}_i}
(-C_0(\tilde{C}_j,\sfp,\tilde{N}_i))\right] . \nonumber \\
&&+C_fg_s^2[4C_{24}(\sg,\sfm,\sfp)+B_1(\sg,\sfp)+B_1(\sg,\sfm)], \label{5}
\eea
where $C_q=4/3$ and $C_{\ell}=0$.
The box correction $\delta_f^{(b)}$ is
\bea
(4\pi)^2\delta_f^{(b)}&=&
-4g_Z^2\sum_{i,j}\left[ g_{NiL}^{(e)}g_{NiL}^{(f_2)*}|U_{j1}|^2
m_W^2D_{27}(\sfm,\se_L,\tilde{C}_j,\tilde{N}_i) \right. \nonumber\\
&&\left. +g_{NiL}^{(\nu)*}g_{NiL}^{(f_1)}|V_{j1}|^2
m_W^2D_{27}(\sfp,\snu_e,\tilde{C}_j,\tilde{N}_i)\right] \nonumber\\
&&-2g_Z^2\sum_{i,j}\left[ g_{NiL}^{(f_2)*}g_{NiL}^{(\nu)*}U_{j1}V_{j1}
m_{\tilde{C}_j}m_{\tilde{N}_i}m_W^2D_0(\sfm,\snu_e,\tilde{C}_j,\tilde{N}_i)
\right. \nonumber\\
&&\left. +g_{NiL}^{(e)}g_{NiL}^{(f_1)}U_{j1}^*V_{j1}^*
m_{\tilde{C}_j}m_{\tilde{N}_i}m_W^2D_0(\sfp,\se_L,\tilde{C}_j,\tilde{N}_i)
\right] . \label{6}
\eea
Here the couplings of charginos $\tilde{C}_j$ and
neutralinos $\tilde{N}_i$ are given by
\bea
g_{NiL}^{(f)}&=&I_{3f}N_{i2}\cos\theta_W+(Q_f-I_{3f})N_{i1}\sin\theta_W ,
\nonumber \\
l_{ij}^{0-}&=&N_{i2}U_{j1}^*+\frac{1}{\sqrt{2}}N_{i3}U_{j2}^* ,\nonumber \\
r_{ij}^{0-}&=&N_{i2}^*V_{j1}-\frac{1}{\sqrt{2}}N_{i4}^*V_{j2} ,\label{7}
\eea
and $g_Z=g/\cos\theta_W$.
$U$, $V$, $N$ are the mixing matrices for charginos and neutralinos \cite{gh}.
The masses of external fermions are ignored in eqs.(\ref{5},\ref{6}).
We adopt the notation of ref.\cite{hhkm} for the
$B,C,D$ functions \cite{pv}.
All 4-momenta in these functions are set to zero.
The results (\ref{5},\ref{6}) generalize the $f=\ell$ case
in refs.\cite{sola,chankow}
and the $\tan\beta=1$ case in ref.\cite{barbieri1}.

The quark-lepton universality violation (\ref{3}) is now expressed as
\bea
\delta_{q\ell}\equiv
\frac{\delta G_q}{G_q}-\frac{\delta G_{\mu}}{G_{\mu}}&\equiv &
(|V_{ud}|^2+|V_{us}|^2+|V_{ub}|^2)^{\frac{1}{2}}-1 \nonumber \\
&=&\delta_q-\delta_{\ell} \nonumber \\
&=&(\delta_q^{(v)}+\delta_{\ell}^{(v)}+\delta_q^{(b)})
-(2\delta_{\ell}^{(v)}+\delta_{\ell}^{(b)}) \nonumber \\
&=&(\delta_q^{(v)}+\delta_q^{(b)})
-(\delta_{\ell}^{(v)}+\delta_{\ell}^{(b)}) \nonumber \\
&=&-0.0009\pm 0.0008. \label{8}
\eea

In the following numerical estimates of $\delta_{q\ell}$,
we assume the generation independence of the sfermion masses,
so that the relevant SUSY particle masses and couplings
are parameterized in terms of the seven variables
($M_{\tilde{Q}}$, $M_{\tilde{L}}$, $M_2$, $M_1$, $\mu$, $\tan\beta$,
$M_{\sg}$) \cite{nilles,mssm,gh}.
We further impose the following mass relations suggested by the minimal
supergravity model with grand unification \cite{nilles}
\beq
M_1=\frac{5}{3}M_2\tan^2\theta_W,\;\;\;
M_{\sg}=\frac{g_s^2}{g^2}M_2,\;\;\;
M_{\tilde{Q}}^2=M_{\tilde{L}}^2+9M_2^2,
\label{9}
\eeq
to reduce the number of independent parameters to four.
The standard model parameters are set as
$m_W=80.24$GeV, $\sin^2\theta_W=0.2312$, $\alpha(m_Z)=1/128.72$ and
$\alpha_s(m_Z)=0.12$.

In Fig.1, the quark-lepton universality violation
$\delta_{q\ell}$ is shown in the ($M_2$, $\mu$) plane for several values of
($m_{\snu}$, $\tan\beta$). The solid lines are contours for
constant $\delta_{q\ell}$'s.
The regions below the thick solid lines ($\delta_{q\ell}=-0.0001$)
are consistent with the 1-$\sigma$
universality violation (\ref{3}). The regions below the thick dashed lines
are excluded by LEP-I experiments. Therefore the regions under the
thick solid lines and above the thick dashed lines are favored
by the present data.
For the ($m_{\snu}=200$GeV, $\tan\beta=2$) case (Fig.1c), however,
the favorable regions lie in the $|\mu|>500$GeV region,
outside of the frame of the figure.
As seen in the figure, the SUSY parameters which
satisfy the universality violation (\ref{3}) and the LEP-I bound tend to lie
in the $M_2\simlt|\mu|$ region, where the lighter chargino and neutralinos are
gaugino-like. On the other hand, when
$M_2\simgt|\mu|$, $\delta_{q\ell}$ tends to be positive and disfavors
the negative deviation (\ref{3}). We find that the 1-$\sigma$ allowed
region in the ($M_2$, $\mu$) plane reduces with increasing $m_{\snu}$, while
the $\tan\beta$ dependence is not significant. For example, we find that
the favored regions for the ($m_{\snu}=100$GeV, $\tan\beta=10$) case are
similar to those for the ($m_{\snu}=100$GeV, $\tan\beta=2$) case (Fig.1b).
Note, however, that under the LEP-I kinematical constraints,
the SUSY contribution to $\delta_{q\ell}$ cannot reach its present
central value of eq.(\ref{8}), $-0.0009$.

In Fig.2, the 1-$\sigma$ allowed region of masses of the
sneutrino \snu and the lighter chargino $\tilde{C}_1$ from
the quark-lepton universality violation (\ref{3}) is shown.
Although the allowed regions in Fig.1 are unbounded in the
directions of $|\mu|\to\infty$, $m_{\tilde{C}_1}$ is bounded from above.
The 1-$\sigma$ (67\% C.L.) upper bounds are roughly
$m_{\snu}<$220GeV and $m_{\tilde{C}_1}<$600GeV, respectively.
Therefore, the 1-$\sigma$ deviation (\ref{3}) from the
quark-lepton universality tends  to favor light sleptons
and relatively light chargino and neutralinos with significant
gaugino components.
It is interesting that the upper bound of $m_{\snu}$
increases with increasing $m_{\tilde{C}_1}$
for $50{\rm GeV}<m_{\tilde{C}_1}<100{\rm GeV}$,
similar to the case of $\delta_{\ell}$ that has been
studied in ref.\cite{chankow}.

Finally, we examine implications of the sign of $\delta_{q\ell}$ for the
SUSY model.
As seen in Fig.1, $\delta_{q\ell}$ takes both signs,
contrary to the result of ref.\cite{barbieri1} where only cases with
very light sfermions ($M_{\tilde{L}}<m_Z/2$, $M_{\tilde{Q}}<m_Z$)
were studied and
only negative $\delta_{q\ell}$ was found.
In fact, it is a cancellation between the vertex and box corrections
that causes the sign change and the non-monotonically decreasing
behavior of $\delta_{q\ell}$ as mentioned above.
In Fig.3, each term in the third line of eq.(\ref{8})
is shown with their sums.
Individual corrections $\delta^{(v)}$'s, $\delta^{(b)}$'s take
definite signs, namely $\delta_{\ell,q}^{(v)}<0<\delta_{\ell,q}^{(b)}$,
while the sum $\delta_{\ell}$ takes both signs.
We can see that the universality violation $\delta_{q\ell}$ is basically
determined by $\delta_{\ell}^{(v)}+\delta_{\ell}^{(b)}$.
The reason is that the squarks are heavier
than the sleptons in our assumption (\ref{9}). We find that
if we set $M_{\tilde{Q}}=M_{\tilde{L}}$ instead of (\ref{9}),
the magnitude of $|\delta_{q\ell}|$ is much reduced.
We also find that the gluino contribution to $\delta_{q\ell}$ is completely
negligible, less than ${\cal O}(10^{-6})$, for our parameter choice.

In conclusion, we have studied the SUSY contribution to the
quark-lepton universality violation in low-energy charged current
interactions, for the whole parameter space of the MSSM.
We have shown that the 1-$\sigma$
universality violation (\ref{3}) in recent data may be
a signal of the light sleptons and relatively light
chargino and neutralinos with significant gaugino components.
We have also briefly discussed the sign of the SUSY contribution
to $\delta_{q\ell}$.
Although it is a hard task to reduce the uncertainty of the CKM matrix
elements in eq.(\ref{2}), further improvements in the
determination of the matrix elements $(V_{ud},V_{us})$ will give
important informations on the SUSY particles.

\section*{\large Acknowledgements}
\pr
We thank M.~Drees for calling our attention to the work of ref.\cite{leut}.
The work of Y.~Y. is supported in part by the JSPS Fellowships
and the Grant-in-Aid for Scientific Research from the Ministry
of Education, Science and Culture of Japan No. 07-1923.


\def\PL #1 #2 #3 {Phys.~Lett. {\bf#1}, #2 (#3) }
\def\NP #1 #2 #3 {Nucl.~Phys. {\bf#1}, #2 (#3) }
\def\ZP #1 #2 #3 {Z.~Phys. {\bf#1}, #2 (#3) }
\def\PR #1 #2 #3 {Phys.~Rev. {\bf#1}, #2 (#3) }
\def\PP #1 #2 #3 {Phys.~Rep. {\bf#1}, #2 (#3) }
\def\PRL #1 #2 #3 {Phys.~Rev.~Lett. {\bf#1}, #2 (#3) }
\def\PTP #1 #2 #3 {Prog.~Theor.~Phys. {\bf#1}, #2 (#3) }
\def\ib #1 #2 #3 {{\it ibid.} {\bf#1}, #2 (#3) }
\def\etal {{\it et al}.}
\def\eg {{\it e.g}.}
\def\ie {{\it i.e}.}


\vspace{20mm}
\section*{Figure Captions}
\renewcommand{\labelenumi}{Fig.\arabic{enumi}}
\begin{enumerate}

\vspace{6mm}
\item
The SUSY contribution to the quark-lepton universality violation
parameter $\delta_{q\ell}$ in the $(M_2, \mu)$ plane for
$(m_{\snu}({\rm GeV}),\tan\beta)=$(50,2)(a), (100,2)(b),
(200,2)(c) and (100,1)(d).
The SUSY contribution explains the universality violation (\ref{3})
at the 1-$\sigma$ level in regions below the thick solid lines.
In Fig.1c, the allowed regions are outside of the frame.
The regions below the thick dashed lines are excluded by LEP-I
experiments.

\vspace{6mm}
\item
The 1-$\sigma$ allowed region of ($m_{\tilde{C}_1}$, $m_{\snu}$)
for explaining the universality violation (3) as the SUSY contribution
for $\tan\beta=(1,2,10)$.

\vspace{6mm}
\item
$\delta_q^{(v)}$, $\delta_{\ell}^{(v)}$, $\delta_q^{(b)}$,
$\delta_{\ell}^{(b)}$ and
their sums as functions
of $\mu$ for $M_2=200$GeV, $m_{\snu}=100$GeV and $\tan\beta=2$.

\end{enumerate}


\newpage
\begin{thebibliography}{99}

\bibitem{towner}
I.S. Towner, \PL B333 13 1994 .

\bibitem{sirlin2}
A. Sirlin, talk at the Ringberg Workshop on ``Perspectives for
electroweak interactions in $e^+e^-$ collisions'', February 1995,
hep-ph/9504385 .

\bibitem{rpp}
Particle Data Group, \PR D50 1173 1994 .

\bibitem{sirlin}
A. Sirlin, Rev. Mod. Phys. {\bf 50}, 971 (1980);\NP B196 83 1982 ;\\
W.J. Marciano and A. Sirlin, \PRL 56 22 1985 ;\\
A. Sirlin and R. Zucchini, \PRL 57 1994 1986 ;\\
A. Sirlin, \PR D35 3423 1987 ;\\
W. Jaus and G. Rasche, \PR D35 3420 1987 .

\bibitem{leut}
H. Leutwyler and M. Ross, \ZP C25 91 1984 .

\bibitem{nilles}
H. P. Nilles, \PP 110 1 1984 .

\bibitem{mssm}
H.E. Haber and G.L. Kane, \PP 117 75 1985 ;\\
R. Barbieri, Riv. Nuovo Cimento {\bf 11}, 1 (1988) .

\bibitem{barbieri1}
R. Barbieri, C. Bouchiat, A. Georges and P. Le Doussal, \PL 156B 348 1985 ;
\NP B269 253 1986 .

\bibitem{barbieri2}
R. Barbieri, M. Frigeni and F. Caravaglios, \PL B279 169 1992 .

\bibitem{hhkm}
K.\,Hagiwara, D.\,Haidt, C.S.\,Kim and S.\,Matsumoto, \ZP C64 559 1994 .

\bibitem{towner2}
I. S. Towner, \NP A540 478 1992 ;\\
F. C. Barker, B. A. Brown, W. Jaus and G. Rasche,
\NP A540 501 1992 .

\bibitem{gh}
J. F. Gunion and H. E. Haber, \NP B272 1 1986 ; \ib B402 567 1993 (E).

\bibitem{pv}
G. 't Hooft and M. Veltman, \NP B153 365 1979 ;\\
G. Passarino and M. Veltman, \NP B160 151 1979 .

\bibitem{sola}
J.A. Grifols and J. Sol\`a, \NP B253 47 1985 .

\bibitem{chankow}
P.H. Chankowski \etal , \NP B417 101 1994 .

\end{thebibliography}
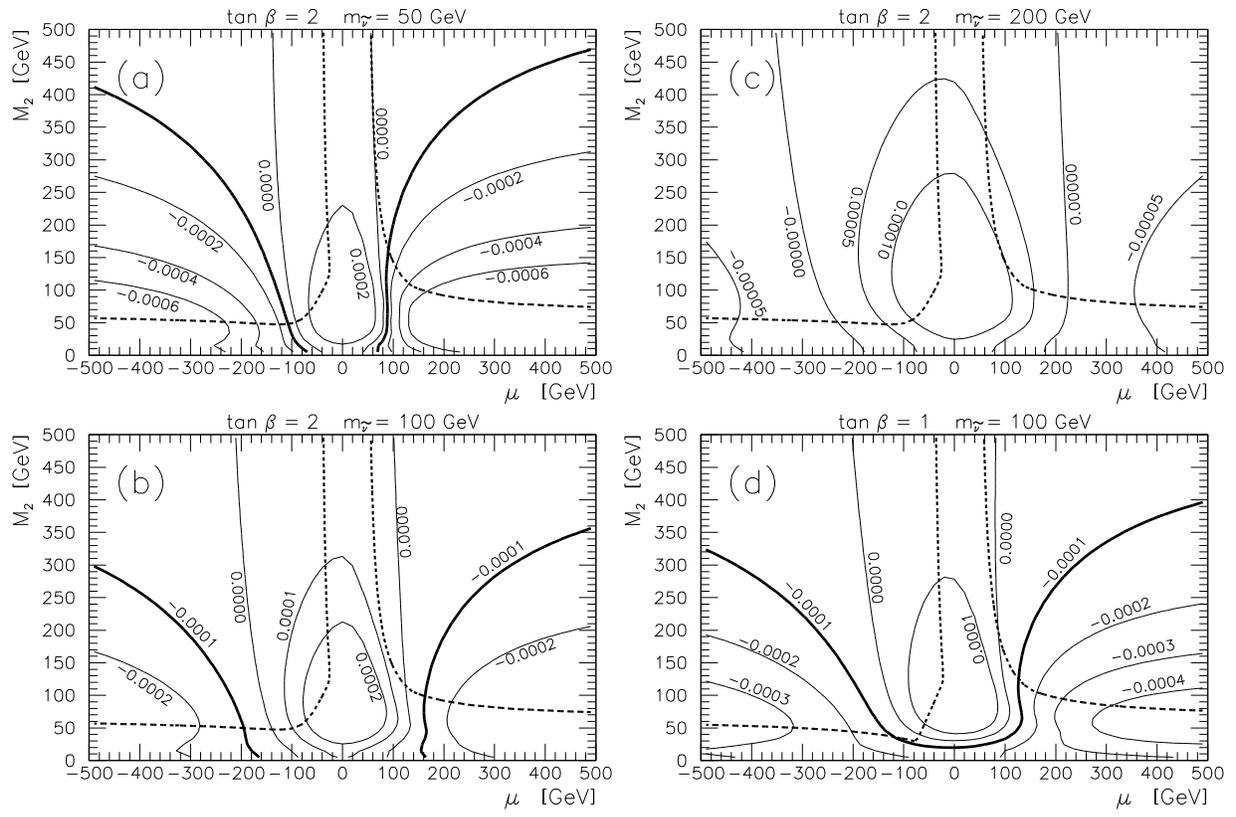
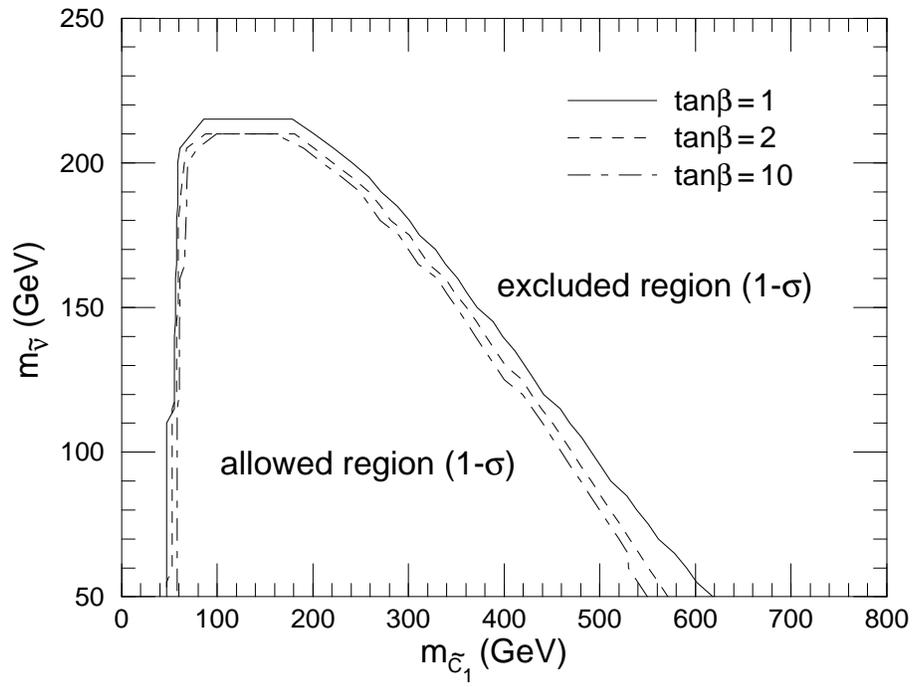
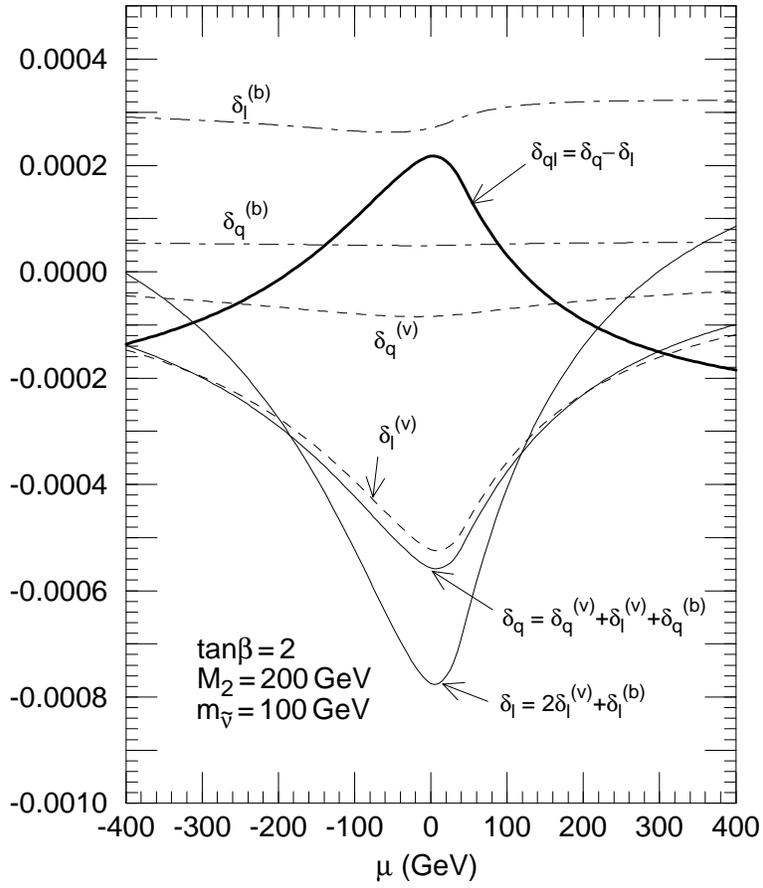
\end{document}